\documentclass[namedreferences]{solarphysics}
\usepackage[hyperref, optionalrh, natbib]{spr-sola-addons} 
\usepackage{makecell}	   
\usepackage{rotating}		
\usepackage{pdflscape}		
\usepackage{longtable}		
\usepackage{booktabs}		
\usepackage{caption}		
\usepackage{tabularx}		
\usepackage{graphicx}                    
\usepackage{color}   
\usepackage{breakurl}                    

\newcommand{\degree}{^{\circ}}

\newcommand{\Rsun}{R$_{\odot}$}
\newcommand{\eg}{\textit{e.g.~}}
\newcommand{\ie}{\textit{i.e.~}}
\newcommand{\insitu}{{\it in situ~}}

\begin{document}

\begin{article}

\begin{opening}

\title{Low-frequency type II radio detections and coronagraph data to describe and forecast the propagation of 71 CMEs/shocks}

%
\author{H.~\surname{Cremades}$^{1,2}$\sep
        F. A.~\surname{Iglesias}$^{1,*}$\sep
        O. C.~\surname{St. Cyr}$^{3}$\sep
        H.~\surname{Xie}$^{3,4}$\sep
        M. L.~\surname{Kaiser}$^{3}$\sep
        N.~\surname{Gopalswamy}$^{3}$
       }

%
\runningauthor{Cremades et al.}
\runningtitle{Low-frequency type II emissions and CME propagation}

%
\institute{$^{1}$ Universidad Tecnol\'ogica Nacional - Facultad Regional Mendoza, Mendoza, Argentina\\
                  email: \url{hebe.cremades@frm.utn.edu.ar}\\ 
		     $^{2}$ Consejo Nacional de Investigaciones Cient\'ificas y T\'ecnicas, Argentina\\
           $^{3}$ NASA Goddard Space Flight Center, Greenbelt, Maryland, USA\\
           $^{4}$ Department of Physics, Catholic University of America, Washington DC, USA\\
           $^{*}$ Now at Max Planck Institut f\"ur Sonnensystemforschung, G\"ottingen, Germany\\ 
          }

\begin{abstract}
The vulnerability of technology on which present society relies demands that a solar event, its time of arrival at Earth, and its degree of geoeffectiveness be promptly forecasted. Motivated by improving predictions of arrival times at Earth of shocks driven by coronal mass ejections (CMEs), we have analyzed 71 Earth-directed events in different stages of their propagation. The study is primarily based on approximated locations of interplanetary (IP) shocks derived from type II radio emissions detected by the  {\it Wind}/WAVES experiment during 1997-2007. Distance-time diagrams resulting from the combination of white-light corona, IP type II radio, and \insitu data lead to the formulation of descriptive profiles of each CME's journey toward Earth. Furthermore, two different methods to track and predict the location of CME-driven IP shocks are presented. The linear method, solely based on {\it Wind}/WAVES data, arises after key modifications to a pre-existing technique that linearly projects the drifting low-frequency type II emissions to 1 AU. This upgraded method improves forecasts of shock arrival time by almost 50\%. The second predictive method is proposed on the basis of information derived from the descriptive profiles, and relies on a single CME height-time point and on low-frequency type II radio emissions to obtain an approximate value of the shock arrival time at Earth.  In addition, we discuss results on CME-radio emission associations, characteristics of IP propagation, and the relative success of the forecasting methods.

\end{abstract}
%
\end{opening}


%
\section{Introduction}
\label{s:introduction} 

Certainly Coronal Mass Ejections (CMEs) are one of the most impressive consequences of solar dynamics. They have acquired growing importance in the field of space weather forecasting, mainly motivated by their recognized capability to interact with Earth's magnetosphere and ultimately trigger geomagnetic storms, with possible harmful consequences for various human technologies. CMEs can drive extensive magnetohydrodynamic (MHD) shocks, as they travel in the interplanetary (IP) medium carrying out vast amounts of plasma and magnetic fields at velocities that can surpass 2000 km s$^{-1}$ (\eg \opencite{Yashiro-etal2004}). These shocks excite electrons, which in turn produce a radio emission at the local plasma frequency, related with the local electron density through $f$[kHz]$=9\sqrt{n_e}$ [cm$^{-3}$], and/or its first harmonic (\eg \opencite{Reiner-etal1997}). As the shock encounters regions of lower density, the frequency of the emission decreases giving rise to a slowly drifting radio emission called type II radio burst (TII). Metric TII bursts may start at around 400 MHz close to the Sun, while kilometric emissions may reach down to $20$ kHz at L1 spacecraft. Due to the Earth's ionosphere filtering effect, the detection of these longer wavelengths is only possible by means of space-borne instruments. 

The first \insitu observation of the source region of a TII radio emission by \citet{Bale-etal1999} confirmed previous analyses \citep{Reiner-etal1997, Reiner-etal1998a, Reiner-etal1998b} that indicated that TII emissions are generated in the upstream region of CME-driven shocks (hereafter CMEs/shocks). Multiple studies on the source regions of TII emissions have followed. For instance, \citet{Knock-etal2001, Knock-etal2003} and \citet{Knock-Cairns2005} analyze the TII emissions originating at shocks moving through various environments, such as a quiet corona and solar wind, coronal loops, CIRs, and preexisting CMEs, on the basis of a theoretical model that predicts the source of the TII radio emission to lie in the foreshock region upstream of a MHD shock front. \citet{Reiner-etal1998a} and \citet{Gopalswamy-etal2001a} revealed the importance of upstream plasma conditions on the detected spectra through correlations between pre-existing plasma structures and changes in emission levels, while \citet{Cho-etal2011} identified the CME nose and CME-streamer interaction as the sites of the multiple TII emissions analyzed by them.

The capability of TII emissions to track CMEs/shocks in the IP medium has been exploited by previous studies focusing on few specific events (\eg \opencite{Pinter1982, Smart-Shea1985, Pinter-Dryer1990}; \opencite{Reiner-etal1998a, Dulk-etal1999, Leblanc-etal2001, Hoang-etal2007}), and by studies combining information extracted from white-light coronagraph images, low-frequency radio spectra, \insitu spacecraft detections, and/or interplanetary scincilation (IPS) data (\opencite{Pohjolainen-etal2007}; \opencite{Cho-etal2007, Feng-etal2009}; \opencite{GonzalezE-AguilarR2009, Bisi-etal2010, Liu-etal2013}; \opencite{Iju-etal2013}). The first joint analysis of white-light, radio, and \insitu observations on a statistically relevant number of cases is that of \citet{Reiner-etal2007}. They describe the propagation profiles of 42 CMEs/shocks occurred during solar cycle 23 by assuming a constant deceleration up to a certain distance, congruent with the \insitu shock arrival time and speed. Their study does not take into account kilometric type II emissions, only evident in the dynamic spectra of the {\it Wind}/WAVES {\it Thermal Noise Receiver} (TNR: see next Section). The kilometric range of TII emissions (kmTII: 300 -- 30 kHz) is of particular interest for the present study, because the distances at which these emissions take place ($\sim$ 20 -- 170 \Rsun) are favourable to elaborate forecasts of shock arrival time (SAT). 

Aside from their usefulness to track and describe the propagation of the shocks driven by CMEs, TII emissions have also been employed to predict the arrival time of their associated shocks at Earth's geospace.  For an overview of TII emissions-based models aimed at forecasting arrival time of shocks at Earth, see the review by \citet{Pick-Vilmer2008}. Some good proxies were also obtained from empirical models based on or combined with other data sets, such as white-light coronal observations (\eg \opencite{Gopalswamy-etal2000, Gopalswamy-etal2001b, Smith-etal2003, Michalek-etal2004, Schwenn-etal2005}), ground-based interplanetary scintillation measurements \citep{Manoharan2006}, solar energetic particles \citep{Qin-etal2009}, novel images from the {\it Heliospheric Imager} (HI: \opencite{Eyles-etal2009}) instruments \citep{Moestl-etal2013, Mishra-etal2013} aboard the {\it Solar-Terrestrial Relations Observatory} (STEREO: \opencite{Kaiser-etal2008}). Additional research comparing ENLIL model \citep{Odstrcil-etal2005} runs for 16 events with the kmTII-based technique described below was reported by \citet{Xie-etal2013}.

The empirical technique based on kmTII emissions by \citet{Cremades-etal2007} (hereafter CSK2007) was conceived to obtain proxies of SAT at Earth. They analyzed distinct data sources for both radio and \insitu observations covering the years 1997 to 2004. For the 92 matched pairs of kmTII emissions and MHD shocks, they derived approximate location and radial speed of the MHD shock at the times of the radio emission. Estimate values of SAT at Earth were then obtained by assuming constant speed.  These SAT values were, however, highly conditioned to the electron density value at 1 AU required by the density model, while in the dynamic spectrum images only two patches of radio emissions were allowed by the selection procedure. These issues are addressed in Section~\ref{s:profiles} and their impact on the prediction technique later in Section \ref{s:linearpred}. Furthermore, an approach that departs from the CSK2007 dataset and introduces information of the associated coronal counterparts helps to describe the propagation of the analyzed events from Sun to Earth (Section \ref{s:profiles}). The outcome provides information on the characteristics of the dataset (Section~\ref{s:characteristics}) and stimulates the outline of another predictive version of the technique, presented in Section~\ref{s:2ndopred}.

\section{Data Sets}\label{s:data} 

A vast amount of data and catalogs were employed to inspect different propagation stages of CMEs/shocks during the years 1997-2007. The starting point of the study is the CSK2007 list of kmTII-\insitu shock pairs, introduced in the following Subsection, while CME counterparts were associated to the IP pairs of events as described in Subsection~\ref{ss:corona}.

\subsection{Radio and \insitu Data}\label{ss:radio}

All radio data used in this study was provided by the {\it Radio and Plasma Wave} (WAVES) experiment on the {\it Wind} spacecraft \citep{Bougeret-etal1995}. WAVES detects radio emissions in three different spectral ranges by means of three receivers: {\it Radio Receiver Band 2} (RAD2: 13.825-1.075 MHz), {\it Radio Receiver Band 1} (RAD1: 1040-20 kHz), and {\it Thermal Noise Receiver} (TNR: 256-4 kHz). Solar kilometric emissions are filtered by Earth's atmosphere, thus the only way to detect them is by means of space-based instrumentation. As mentioned previously, one of the peculiarities of the present study is that it relies on TNR kilometric emissions, given TNR's much better spectral resolution in most of the kilometric frequency range in comparison with that of RAD1, although the latter does also cover the kilometric wavelength range. TII emissions occurring at or extending to kilometric frequency ranges, i.e. lower than 300 kHz, were extracted from the {\it Wind}/WAVES Type II radio bursts list maintained by M. L. Kaiser (\url{http://www-lep.gsfc.nasa.gov/waves/data_products.html}). From 1997 to 2007, a total of 181 kmTII radio emissions were found. 

MHD shocks detected at L1 by the {\it Advanced Composition Explorer} (ACE: \opencite{Stone-etal1998}) and {\it Wind} spacecraft were obtained from the IP shocks catalogues developed by: D.B. Berdichevsky et al. (\url{http://pwg.gsfc.nasa.gov/wind/current_listIPS.htm}; Wind), the Space Plasma Group at the Massachusetts Institute of Technology - now available at the Harvard-Smithsonian Center for Astrophysics (\url{http://www.cfa.harvard.edu/shocks/}; Wind and ACE); and the Experimental Space Plasma Group at the University of New Hampshire (\url{http://www.ssg.sr.unh.edu/mag/ace/ACElists/obs_list.html}; ACE). After cross-referencing information from all of these catalogs, 333 forward shocks were identified during the years 1997 to 2007. In principle, a shock is considered to be potentially associated to a kmTII emission if it occurs within 72 h after the appearance of the radio emission. 
If no candidate shock was found within that time span, the following time interval was investigated. Only in three cases (out of the analyzed 71 events) a shock was detected after three days, in particular after 79 h (event \#3 --see Table~\ref{tbl:associations}), 93 h (\#4), and 82 h (\#10).

\subsection{Coronal Data}\label{ss:corona}

Backtracking of the kmTII-shock IP features to CMEs in the solar corona involved white-light data provided by the {\it Large Angle Spectroscopic Coronagraph} (LASCO: \opencite{Brueckner-etal1995}) onboard the {\it Solar and Heliospheric Observatory} (SOHO: \opencite{Domingo-etal1995}).  Firstly, the existence of CME events in agreement with temporal and spatial considerations was ascertained by inspection of  the CDAW SOHO/LASCO CME Catalog (\url{http://cdaw.gsfc.nasa.gov/CME_list}; \opencite{Yashiro-etal2004}). Preliminary associations were subsequently compared and verified with the aid of CME-interplanetary CME (ICME) lists, namely: the Richardson \& Cane Near-Earth ICMEs in 1996-2007 (\url{http://www.ssg.sr.unh.edu/mag/ace/ACElists/ICMEtable.html}; henceforth RC ICME list), the list of associations during 1997-2005 in Table 1 of \citet{Gopalswamy-etal2005}, and R. Schwenn's list of CME-ICME associations during 1997-2001 (personal communication).

\begin{figure} 
\centering
\includegraphics[width=0.75\textwidth]{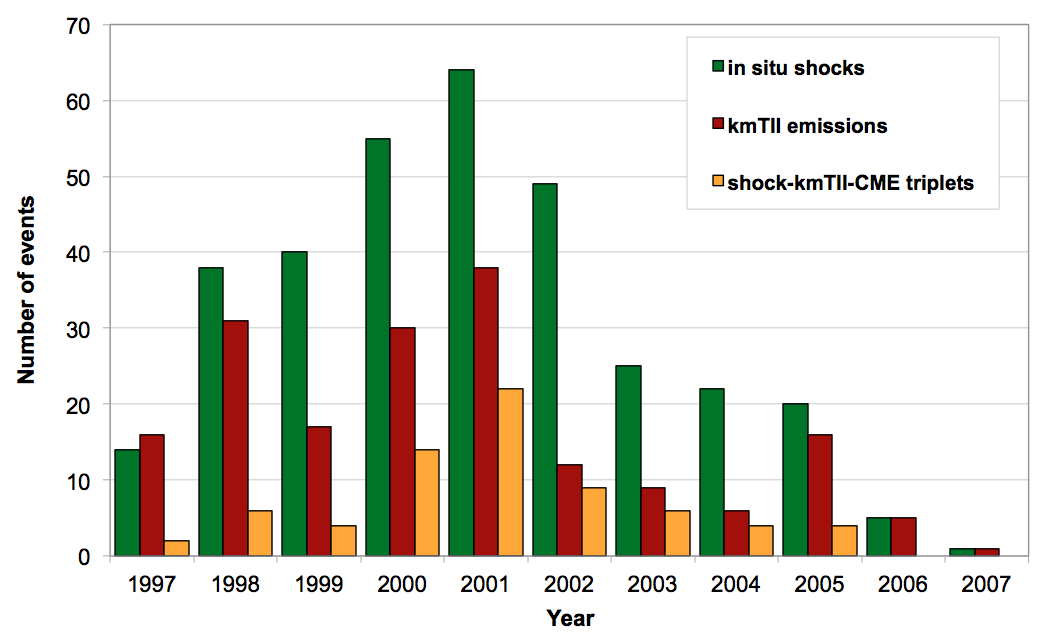}
\caption{Yearly frequency of \insitu forward shocks (green bars), kmTII radio emissions (red bars), and kmTII-shock pairs that could be associated to a CME (orange bars).}
\label{fig:yearlyfreq}
\end{figure}

Thirteen kmTII-shock pairs from the original CSK2007 list with 94 events were discarded, either because their speed was implausibly slow, because the CME onset time and the backtracking of the kmTII distance-time profile were incompatible, or because the CME source location and/or angular width was inconsistent with an Earth-directed radio emission. To the remaining 81 events, four corresponding to the year 2005 were added, while 14 kmTII-shock pairs occurred after or during SOHO/LASCO data gaps. These issues reduced the data set to 71 events. We also note that none of the kmTII-shock pairs during 2006 and 2007 could be associated to a CME. Figure~\ref{fig:yearlyfreq} shows the yearly frequency of forward shocks (green bars) and kmTII radio emissions (red bars), in comparison with the number of kmTII-shock pairs that could be associated to a CME (orange bars). The solar cycle variation is in agreement with the one found by \citet{Gopalswamy-etal2010} during 1996-2006 for associations between \insitu shocks and metric-kilometric TII emissions. The 71 triple associations are listed later in Table~\ref{tbl:associations} of Section \ref{s:characteristics}.

\section{Propagation profiles}\label{s:profiles}
\subsection{Methodology}\label{ss:methodology}

The various stages of propagation were put together to assemble distance-time plots for each of the 71 CME-kmTII-shock events: CME height in the solar corona, IP distance derived from the kmTII emissions, and arrival time of the shock detected \insitu at {\it Wind} or ACE. Height information of CMEs in the solar corona was obtained from the CDAW SOHO/LASCO CME Catalog. Since it provides height values of three-dimensional entities projected in the two-dimensional plane of the sky, these should be interpreted with caution. Projection effects in these coronal height-time points are addressed in the next subsection. 

\begin{figure} 
\centering
\includegraphics[width=0.99\textwidth]{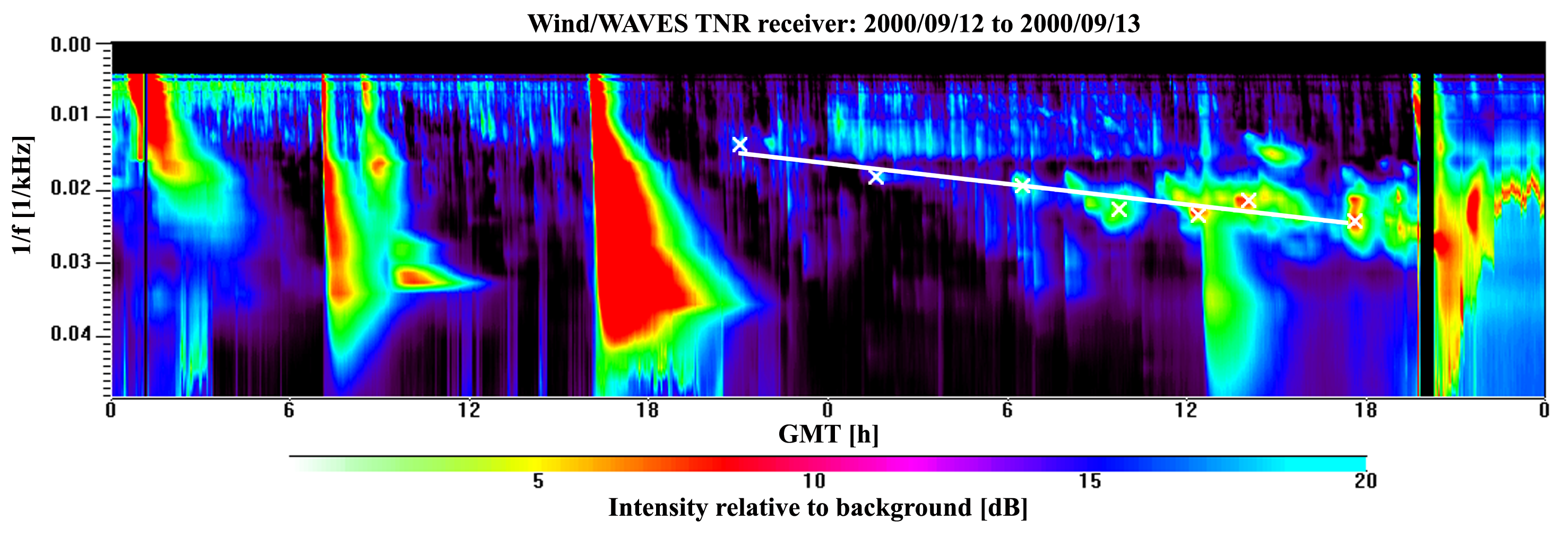}
\caption{KmTII emission detected by {\it Wind}/WAVES TNR on September 13, 2004. The ordinates are plotted in the inverse of the frequency [kHz$^{-1}$], while abscissas represent time [h]. Intensity in dB is colour-coded.}
\label{fig:TIIemission}
\end{figure}

Interplanetary propagation of shocks ahead of CMEs is tracked by means of kmTII emissions discernible in TNR dynamic spectra, on the basis of the relationship between local plasma frequency and density. A typical kmTII emission detected by TNR is displayed in Figure~\ref{fig:TIIemission}, where the vertical axis represents the inverse of the frequency in units of kHz$^{-1}$, and the horizontal axis time in h. Dynamic spectral plots in the 1/f space show drifting radio emissions approximately organized along straight lines, given that 1/f can be assumed to be equivalent to the heliocentric distance R by considering the IP plasma density to roughly vary as 1/R$^{2}$ \citep{Bougeret-etal1984, Reiner-etal1998b}. Naturally, this does not hold true for complex events and when pre-existing structures are present in the IP medium \citep{Knock-Cairns2005}. In Figure~\ref{fig:TIIemission}, white crosses are the points manually selected as representative of the kmTII emission. While CSK2007 took into account only two representative points to obtain the slope of the drifting emission, here we introduce the posibility of selecting several points. This methodology reduces the susceptibility of the slope determination to errors in the selected data points and thus helps to relax the selection process, given that TII emissions are commonly intermittent and of varying bandwidth, in a noisy and contaminated environment. The effect of varying bandwidth is neglected in this approach, which systematically considers the selection of the central point of each kmTII patch.

\begin{figure} 
 \centering
 \includegraphics[width=0.99\textwidth]{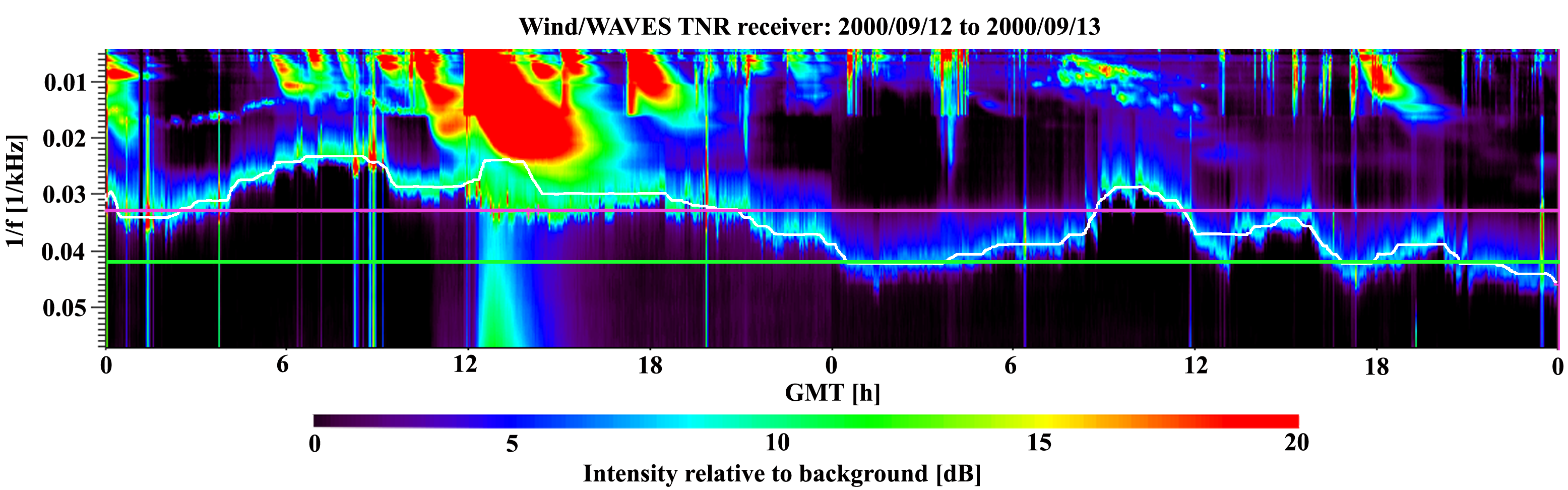}
 \caption{TNR dynamic spectrum during 12-13 September 2000. The white line on top of the plasma frequency line corresponds to the electron density at L1 derived from the CDAWeb data. The mean value during that time period is represented by the pink line at $\approx$31 kHz (11.6 cm$^{-3}$). For comparison, the green line is drawn at the typical value of 24 kHz (7.2 cm$^{-3}$). The considered kmTII emission takes place from 0 to 12 h on 13 September.}
 \label{fig:N0value}
 \end{figure}

TII emission points in the frequency domain are translated to distance from the Sun by means of the relationship between local plasma frequency and density $f$(kHz)$=9\sqrt{n_e}$ (cm$^{-3}$) in combination with the \citet{Leblanc-etal1998} coronal/interplanetary density model. According to this empirical model,  derived from {\it Wind}/WAVES and ground-based radio observations, the electron density $n_e$ decreases with increasing heliospheric distance $r$ in units of \Rsun~as $n_e (r)=\left(3.3\times 10^{5} r^{-2}+4.1\times 10^{6}r^{-4} + 8.0 \times 10^{7} r^{-6} \right) cm^{-3} $. The equation is solved for $r$ by using a globally convergent Newton's method. This model is valid for a density at 1 AU $n_0 = 7.2$ cm$^{-3}$, while for individual bursts with different $n_0$ the model is multiplied by $n_0 / 7.2$. The electron density value at 1 AU, $n_0$, is a crucial input to the model that directly affects the calculation of the shock location in the IP medium. According to \citet{Leblanc-etal1998}, assuming $n_0$ = 7.2 cm$^{-3}$ is a good approximation for all cases when the fluctuation of its real value prevents adopting a better one. This was the criterion used by CSK2007 in most cases. However, this becomes a major source of error due to the wide variation range of $n_0$ (2 to 39 cm$^{-3}$), especially in times of solar maximum. To reduce the uncertainty in $n_0$, a more realistic value was adopted, based on results of a neural network procedure that detects the local plasma density at the spacecraft \citep{Bougeret-etal1995}, whose outcome is available at the Coordinated Data Analysis Web (CDAWeb; \url{http://cdaweb.gsfc.nasa.gov/}). More precisely, the $n_0$ value used to feed the \citet{Leblanc-etal1998} density model was computed as the mean value during the day(s) in which a specific kmTII event was observed. Figure~\ref{fig:N0value} displays the dynamic spectrum corresponding to the kmTII on 13 September 2000. This is an example of the plasma frequency fluctuating at L1, with an average value considerably differing from 7.2 cm$^{-3}$. The white line following the plasma frequency line represents the high resolution values obtained from CDAWeb, after a filtering process that eliminates high frequency noise. The pink horizontal line at $\approx$31 kHz (or 11.6 cm$^{-3}$) stands for the mean value used as input for the density model, while the green line at 24 kHz corresponds to 7.2 cm$^{-3}$. 

\begin{figure}
   \centerline{\hspace*{0.015\textwidth}
               \includegraphics[width=0.49\textwidth,clip=]{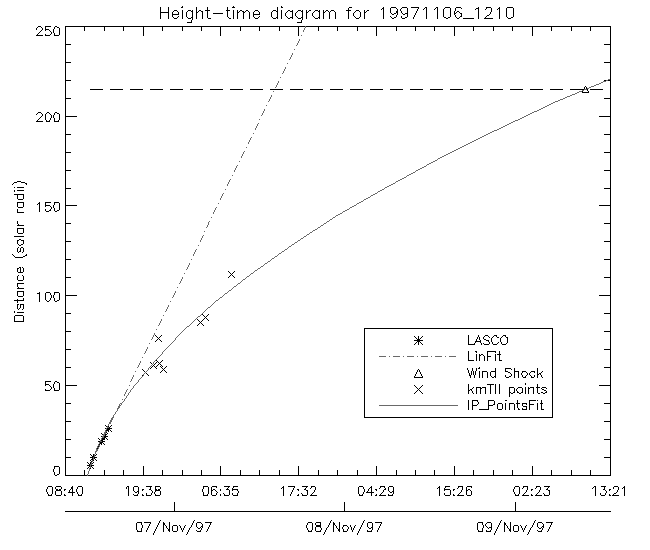}
               \hspace*{-0.03\textwidth}
               \includegraphics[width=0.49\textwidth,clip=]{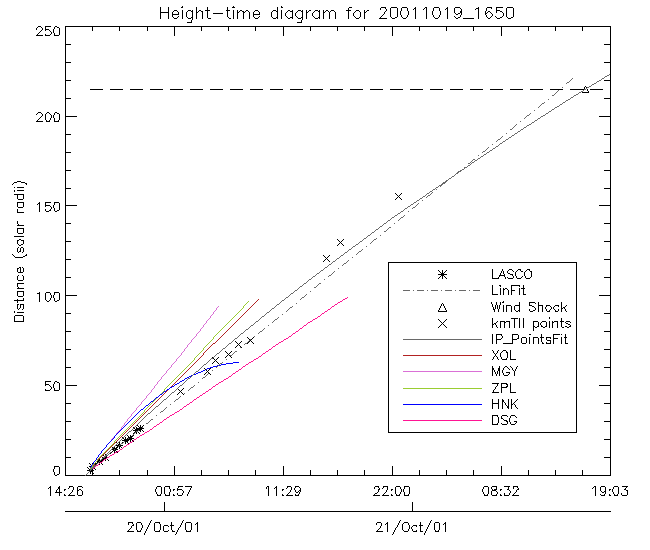}
              }
	\caption{The three stages of CME/shock propagation: CME (asteriscs) and kmTII (crosses) distance-time points, and shock arrival at 1 AU (triangle). The black solid line is the best fit, while the dash-dotted line is a linear fit through the SOHO/LASCO points. The right panel includes several models applied to SOHO/LASCO height-time points to remove projection effects (solid coloured lines, see text).}
   \label{fig:profiles}
   \end{figure}

Interplanetary distance of the CME/shock derived in this way from the kmTII radio emissions, together with the CME height-time measurements from coronal data and the SAT at 1 AU can be combined in single distance-time plots for each event. Figure~\ref{fig:profiles} shows two examples, with the SOHO/LASCO height-time points represented by asterisks, the kmTII-derived distance information by crosses, and the SAT point at 1 AU by a triangle. As mentioned earlier, the SOHO/LASCO height-time points have inherent projection effects. As for the kmTII distance-time points, it is assumed here that the source of the radio emission (namely the shock's emitting parcel) is approximately travelling along the Sun-Earth line, even if the bulk of the ICME propagates off it.

\subsection{Descriptive profiles}\label{ss:descriptive}
After plotting data points together, and to gain understanding on the various propagation profiles exhibited by this set of events from the Sun to 1 AU, distance-time points are fitted to curves.  Equations~\ref{eqn:acc} and~\ref{eqn:dec} are used to represent the accelerated and decelerated cases respectively:

\noindent\begin{tabularx}{\textwidth}{@{}XX}
\begin{equation}
d (t) = a t^{2} +b t +c
\label{eqn:acc}
\end{equation} &
\begin{equation}
d (t) = \sqrt{a t +b}+c
\label{eqn:dec}
\end{equation}
\end{tabularx}

Where $d$ is distance from the Sun, $t$ is time, and $a$, $b$, $c$, are coefficients that arise from the fitting, performed by means of a Levenberg-Marquardt least-squares fit \citep{Markwardt2009}. Naturally, time $t$ and distance $d$ must be positive. The equation for the decelerated case is no other than equation~\ref{eqn:acc} solved for $t$, \ie its inverse, with variable names $t$ and $d$ swapped. Its behaviour is very similar to that of equation~\ref{eqn:acc} with a negative value of $a$, but with a steeper slope close to the Sun, which approximates the sets of data points gathered at different stages of propagation.

The left panel of Figure~\ref{fig:profiles} is an example of a typical decelerated case, while the right panel exhibits a nearly linear propagation profile. In the figure, the black solid line is the henceforth ``descriptive profile", namely the best fit to the kmTII points, connecting only the first appearance of the CME in the coronagraph (to avoid including projection effects) and the shock arrival at 1 AU. The black dash-dotted line is a linear fit through the SOHO/LASCO height-time points, projected to 1 AU.

The right panel of Figure~\ref{fig:profiles} includes, in addition, several models that have been applied to the SOHO/LASCO height-time points to remove projection effects from a nearly Earth-directed event. There are two reasons why it is not possible to straightforward compare the LASCO height-time points with the kmTII and shock arrival points: (i) the SOHO/LASCO points are a projection in the plane of the sky of an approximately Earthward-traveling event, and (ii) most often it is the projected height of the leading edge, and not of the shock, what is measured in coronagraph images. In an attempt to account for the effects of (i), the propagation profiles corrected by several methods (solid coloured lines in Figure~\ref{fig:profiles}, right panel) have been compared with the descriptive profile (black solid line). The considered methods are DSG \citep{DalLago-etal2003}, ZPL \citep{Zhao-etal2002}, MGY \citep{Michalek-etal2003}, XOL \citep{Xie-etal2004}, and HNK \citep{Howard-etal2008}. The DSG is based on the concepts of radial and expansion speed, and its application is straightforward to all events. Radial values of speed derived from the ZPL, MGY, and XOL cone models were used only when available in \citet{Xie-etal2006}, accounting for 17 events in common --enough for the purposes of this study.  The HNK makes use of the CME source region location and 3D aspects of its trajectory, and corrected values of speed and acceleration where provided for 24 of the events here analyzed by D. Nandy (private communication, 2008).

The right panel of Figure \ref{fig:profiles} is one of the 12 cases for which values corrected for projection effects by all methods were available. It is evident from the figure that none of the corrected speed profiles (coloured solid lines) lies close to the descriptive profile based on the kmTII points, the first appearance of the CME, and the shock arrival at 1 AU (black solid line). This was a common situation when attempting to assess the performance of these five methods, with some few cases exhibiting a preference randomly, and not consistently, for one or two of them. It must be noted that the DSG, ZPL, MGY, and XOL methods assume a constant velocity, which is nearly true for 19 events according to the analysis of their speed profiles (see next section). HNK, in spite of being a second-order method, in the general case does not approach to the behaviour of the descriptive propagation profile. Although all of these methods are oriented to deduce the radial speed of a CME and not the Earth-directed component ---not necessarily the same--- it is assumed here that this subset of 12 events propagates approximately along the Sun-Earth line, given that their source region coordinates lie within 30$\degree$ of the central meridian, except for two outliers in longitude (see Table~\ref{tbl:associations}).  In addition, it must be taken into account that even if these models were successful in removing projection effects, their validity is limited to coronal heights, since propagation conditions in the IP medium may drastically differ due to inhomogeneities in the ambient solar wind (\eg \opencite{Pohjolainen-etal2007}).

\section{Characteristics of the events under study}\label{s:characteristics} 

The set of 71 CME-kmTII-shock triplets during 1997-2007 is peculiar on its own, given the low rate of found associations (see Figure~\ref{fig:yearlyfreq}). Therefore, it is of particular interest to investigate various properties of these events starting at their source regions at the Sun and in their different stages of propagation. Table~\ref{tbl:associations} summarizes some of the main characteristics compiled for all analyzed events. In all cases an empty cell indicates an absence of information in the corresponding data source. The first column assigns an event number to each entry, while the following three columns refer to the triple associations made: CME start time as reported by the CDAW SOHO/LASCO CME Catalog, start time of the type II emission as informed by the {\it Wind}/WAVES Type II radio bursts list, and shock arrival time at the {\it Wind} spacecraft.

\begin{figure} 
 \centering
 \includegraphics[width=0.99\textwidth]{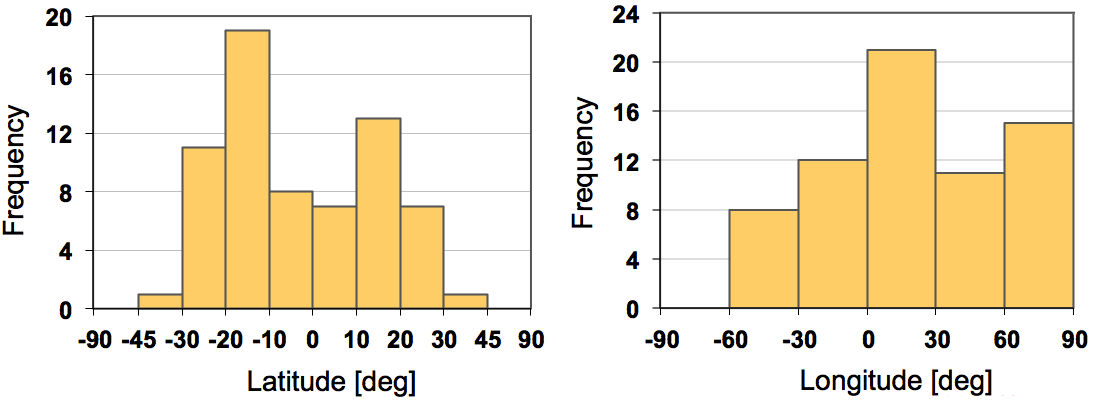}
 \caption{Central latitude (left panel) and longitude (right panel) distributions of the candidate source regions of the CME-kmTII-shock triplets.}
 \label{fig:latlon}
 \end{figure}

The solar origins of the CME events associated to the kmTII/shock pairs were ascertained with the aid of low coronal images provided by the {\it Extreme ultraviolet Imaging Telescope} (EIT: \opencite{Delaboudiniere-etal1995} onboard SOHO. The central heliographic coordinates of the candidate source regions are listed in column 5 of Table~\ref{tbl:associations}. For four events it was not possible to obtain the coordinates, either because there were no SOHO/EIT images, because the source could not be univocally determined, or because it was located behind the western limb. The obtained central coordinates graphed as histograms in Figure~\ref{fig:latlon} show a two-peak latitudinal distribution in agreement with the two activity belts, and a clear preference for western longitudes. \citet{Cliver-etal2004} as well as \citet{Gopalswamy-etal2008} have found a weaker western bias in the sources of CMEs associated with metric and/or decametric-hectometric TIIs. Furthermore, the visibility of solar energetic particles drastically increases towards the west limb in the event of metric and particularly decametric-hectometric TIIs \citep{Cliver-etal2004, Gopalswamy-etal2008}.

As for the CMEs composing the set of analyzed events, 97\% of them turned out to be either full or partial halo CMEs, \ie with angular width greater than 120$\degree$ (see column 6 of Table~\ref{tbl:associations}). The two CMEs that account for the remaining events (\#5 and \#34) are particular cases of CMEs that arose close to disk center, but exhibiting very dim coronal signatures, reason for the scarce measured angular width. Partial halo CMEs account for 20\% of the events. According to the CDAW SOHO/LASCO Catalog classification of full halo CMEs, 38\% of events were of the ``outline asymmetry'' (OA) type, 34\% of the ``brightness asymmetry'' (BA), and 6\% of them were symmetric (S). The large proportion of full halo CMEs of the OA type is in accordance with the predominance of CME source regions with western longitudes.  Despite several CMEs in Table~\ref{tbl:associations} originate close to the west limb, their condition of full halo CMEs indicates their large extent in angular width. Therefore, it is plausible to assume that the kmTII emissions associated to these events originate in the portion of the shock that is travelling in the Earth's direction. Limb full halo CMEs are worth of being considered, given that they have the potential to be geoeffective (\eg, \opencite{Cid-etal2012}).

CME height-time diagrams were built for the 71 events from projected height information provided by the CDAW SOHO/LASCO CME Catalog, {\it i.e.} points are confined to the coronal heights covered by the SOHO/LASCO coronagraphs. Column 7 of Table~\ref{tbl:associations} indicates the type of propagation profile exhibited by each coronal event in the SOHO/LASCO field of view: 43\% appear accelerated, 43\% decelerated, and 13\% linear. Kinematic profiles are considered linear if their average acceleration (second derivative of the distance-time expression of each event's propagation profile) is within $\pm$1.5 m s$^{-2}$. As mentioned before, these values have projection effects attached, and must be interpreted with caution. The behaviour of descriptive profiles obtained as addressed in Section \ref{ss:descriptive}, \ie after including distance-time points derived from the kmTII information and the shock arrival at 1 AU, is listed in column 16 of of Table~\ref{tbl:associations}.  These do not exhibit the same behaviour as in the corona, presumably because of varying conditions in the IP medium that modify the propagation of CMEs/shocks. Statistics of the descriptive profiles yield 39\% of the decelerated type, 34\% accelerated, and 27\% of the linear type, when considering linear events those with acceleration within $\pm$1.5 m s$^{-2}$.

\begin{landscape}
\begin{center}
\tiny
\begin{longtable}{cccccccccccccccccc}
\captionsetup{width=1.61\textwidth}
\caption{CME-kmTII-shock associations and main characteristics of the 71 analyzed events. Column (1): event number. (2) CME start time. (3) TII start time. (4) SAT at {\it Wind}. (5) CME source region heliocentric coordinates. (6) Halo CME type (PH=partial halo, OA=outline asymmetry, BA=brightness asymmetry, S=symmetry; a number indicates the angular width, in degrees, of a non-halo CME). (7) Kinematic profile type exhibited in SOHO/LASCO. (8) Frequency range of the TII [kHz]. (9) TII duration [h]. (10) Adopted values of $n_0$ [cm$^{-3}$]. (11) Speed solely derived from the type II emission. (12) \insitu shock speed [km s$^{-1}$]. (13) Shock transit speed [km s$^{-1}$]. (14) Existence of BDEs in \insitu data. (15) Presence of MC signatures (MFR=only rotation in field direction, (S)=``Schwenn'', (B)=``Berdichevsky {\it et al.}''). (16) Descriptive profile propagation type. (17) Shock initial speed [km s$^{-1}$]. (18) Average acceleration [m s$^{-2}$]. The latter two are derived from the descriptive fit.}
\label{tbl:associations}\\
\toprule
\thead{\#} & \multicolumn{3}{c}{CME-kmTII-shock association} & \multicolumn{3}{c}{CME} & \multicolumn{4}{c}{Type II emission} & \multicolumn{4}{c}{\insitu} &\multicolumn{3}{c}{Descriptive profile}\\(1) & (2) & (3) & (4) & (5) & (6) & (7) & (8) &(9) & (10) & (11) & (12) & (13) & (14) & (15) & (16) & (17) & (18)\\
\cmidrule(lr){2-4} \cmidrule(lr){5-7} \cmidrule(lr){8-11} \cmidrule(lr){12-15} \cmidrule(lr){16-18}
\endfirsthead

\multicolumn{18}{c}%
{{\tablename\ \thetable{} -- continued from previous page}} \\
\thead{\#} & \multicolumn{3}{c}{CME-kmTII-shock association} & \multicolumn{3}{c}{CME} & \multicolumn{4}{c}{Type II emission} & \multicolumn{4}{c}{\insitu} &\multicolumn{3}{c}{Descriptive profile}\\(1) & (2) & (3) & (4) & (5) & (6) & (7) & (8) &(9) & (10) & (11) & (12) & (13) & (14) & (15) & (16) & (17) & (18)\\
\cmidrule(lr){2-4} \cmidrule(lr){5-7} \cmidrule(lr){8-11} \cmidrule(lr){12-15} \cmidrule(lr){16-18}
\endhead

\multicolumn{18}{r}{{Continued on next page}} \\ 
\endfoot

\endlastfoot

1   & 11/04/97 06:10                        & 11/04/97 06:00 & 11/06/97 22:02 & S20W27 & BA                         & Dec & 14000-100 & 23  & 12,55        & 766  &       & 651  & Y    & Y     & Lin         & 533        & 0,95         \\
2   & 11/06/97 12:10                        & 11/06/97 12:20 & 11/09/97 10:03 & S20W53 & OA                         & Dec & 14000-100 & 20  & 13,68        & 562  &       & 596  &      & N (B) & Dec         & 2204       & -7,45        \\
3   & 04/20/98 10:07                        & 04/20/98 10:25 & 04/23/98 17:30 & S17W90 & PH  & Acc & 10000-35  & 44  & 7,23         & 611  & 402   & 524  &      & N (S) & Dec         & 1694       & -4,87        \\
4   & 04/27/98 08:56                        & 04/28/98 00:00 & 05/01/98 21:21 & S17E56 & OA  & Acc & 170-80    & 24  & 5,48         & 328  & 631   & 383  & Y    & Y     & Dec         & 1092       & -2,21        \\
5   & 04/30/98 20:25                 & 05/01/98 12:00 & 05/03/98 17:02 & S17E10 & 45  & Lin & 200-70    & 36  & 12,76        & 547  & 474   & 620  &      & N (S) & Lin         & 623        & -0,07        \\
6   & 06/16/98 18:27                        & 06/16/98 18:20 & 06/18/98 14:25 & S21W90 & PH  & Dec & 12000-50  & 27  & \textbf{7,2} & 876  &       & 945  &      & N (S) & Dec         & 1313       & -3,76        \\
7   & 11/05/98 20:44                        & 11/05/98 22:00 & 11/08/98 04:41 & N18W21 & OA                         & Dec & 5000-50   & 34  & 9,39         & 575  & 645   & 747  & Y    & Y     & Dec         & 978        & -1,99        \\
8   & 11/09/98 18:17                        & 11/11/98 01:00 & 11/12/98 07:18 & N20W18 & PH  & Acc & 150-70    & 17  & 8,27         & 566  & 406   & 690  &      & N (S) & Acc         & 183        & 4,46         \\
9   & 06/30/99 11:54                        & 06/30/99 23:00 & 07/02/99 00:27 & S15W19 & S                          & Dec & 120-70    & 10  & \textbf{7,2} & 946  & 633   & 1135 &      & N     & Dec         & 1525       & -4,74        \\
10  & 07/02/99 16:30                & 07/03/99 04:00 & 07/06/99 14:24 & N11W53 & PH  & Lin & 200-50    & 32  & \textbf{1,5} & 449  & 474   & 448  & N    & MFR   & Lin         & 326        & 0,66         \\
11  & 07/05/99 02:54                        & 07/05/99 18:00 & 07/08/99 04:00 & S25W48 & PH  & Acc & 80-40     & 30  & \textbf{7,2} & 584  &       & 568  &      &       & Lin         & 647        & -0,61        \\
12  & 07/25/99 13:31                        & 07/26/99 01:00 & 07/26/99 23:50 & N38W87 & OA                         & Dec & 150-60    & 15  & \textbf{3}   & 1623 & 390   & 1221 & Y    & N     & Acc         & 1070       & 2,00         \\
13  & 01/28/00 20:12                        & 01/29/00 06:30 & 01/30/00 18:44 & S30W22 & OA                         & Dec & 200-100   & 9   & \textbf{3}   & 991  & 785   & 893  &      & N (S) & Dec         & 1232       & -3,27        \\
14  & 02/08/00 09:30                        & 02/08/00 09:05 & 02/11/00 02:35 & N27E14 & BA                         & Dec & 12000-20  & 65  & 5,79         & 631  & 507   & 642  & Y    & N     & Lin         & 547        & 0,69         \\
15  & 02/10/00 02:30                & 02/11/00 08:45 & 02/11/00 23:34 & N27W12 & BA                         & Acc & 50-23     & 15  & 9,89         & 1150 & 638   & 944  & Y    & Y     & Acc         & 775        & 1,59         \\
16  & 04/04/00 16:33                        & 04/05/00 18:00 & 04/06/00 16:32 & N18W72 & OA                         & Acc & 60-30     & 22  & 7,83         & 1632 & 642   & 875  & Y    & MFR   & Acc         & 263        & 6,46         \\
17  & 05/15/00 16:26                        & 05/15/00 16:45 & 05/17/00 21:35 & S23W71 & PH  & Dec & 4000-40   & 21  & 11,58        & 1473 &       & 782  &      &       & Dec         & 3179       & -14,34       \\
18  & 06/06/00 15:54                        & 06/06/00 15:20 & 06/08/00 08:41 & N21E10 & BA                         & Acc & 14000-40  & 42  & 8,42         & 931  & 868   & 1019 & Y    & N     & Lin         & 1020       & -0,27        \\
19  & 07/14/00 10:54                        & 07/14/00 10:30 & 07/15/00 14:15 & N17W11 & S                          & Dec & 14000-80  & 28  & \textbf{7,2} & 1358 &       & 1519 & Y    & N     & Dec         & 1720       & -4,24        \\
20  & 07/26/00 00:30                        & 07/26/00 09:30 & 07/28/00 06:38 & N08W29 & PH  & Dec & 140-85    & 21  & 12,45        & 681  & 490   & 781  & Y    & Y     & Lin         & 639        & 1,13         \\
21  & 09/04/00 06:06                        & 09/05/00 03:25 & 09/06/00 16:13 & N23W55 & PH  & Acc & 90-50     & 8   & 8,89         & 915  & 538   & 715  &      & N (S) & Acc         & 501        & 1,94         \\
22  & 09/12/00 11:54                        & 09/12/00 12:00 & 09/15/00 04:27 & S17W10 & BA                         & Acc & 14000-60  & 24  & 11,59        & 604  & 377   & 648  &      & N (S) & Lin         & 567        & 0,59         \\
23  & 09/29/00 21:50                        & 09/30/00 13:00 & 10/03/00 01:02 & S09E43 & PH  & Acc & 130-60    & 33  & 8,14         & 705  & 462   & 559  & Y    & Y     & Acc         & 334        & 1,54         \\
24  & 11/01/00 16:26                        & 11/01/00 19:40 & 11/04/00 02:25 & S15E35 & BA                         & Acc & 650-40    & 17  & 8,32         & 526  & 429   & 727  &      &       & Acc         & 383        & 3,11         \\
25  & 11/24/00 05:30                        & 11/24/00 05:10 & 11/26/00 05:32 & N02W17 & BA                         & Acc & 14000-100 & 10  & 10,89        & 1171 & 471   & 875  &      & N (S) & Lin         & 810        & 0,48         \\
26  & 11/24/00 15:30                        & 11/24/00 15:25 & 11/26/00 11:43 & N02W17 & BA                         & Dec & 14000-200 & 7   & 10,89        & 801  & 518   & 947  & Y    & N     & Dec         & 1555       & -5,65        \\
27  & 01/28/01 15:54                        & 01/28/01 15:45 & 01/31/01 08:35 & S06W66 & OA                         & Acc & 14000-200 & 1   & 22,88        & 527  & 488   & 655  &      & N (S) & Acc         & 284        & 3,02         \\
28  & 03/29/01 10:26                & 03/29/01 10:12 & 03/30/01 21:51 & N13W14 & BA                         & Acc & 4000-60   & 20  & 7,67         & 1270 & 529   & 1107 & Y    & MFR   & Lin         & 995        & 1,24         \\
29  & 04/06/01 19:30                        & 04/06/01 19:35 & 04/07/01 17:56 & S21E34 & BA                         & Dec & 14000-60  & 21  & 8,34         & 1521 & 695   & 1065 & Y    & N     & Acc         & 821        & 3,26         \\
30  & 04/09/01 15:54                        & 04/09/01 15:53 & 04/11/01 14:09 & S02W18 & S                          & Lin & 12000-100 & 9   & \textbf{7,2} & 831  & 686   & 917  & Y    & Y     & Dec         & 1164       & -2,73        \\
31  & 04/10/01 05:30                        & 04/10/01 05:24 & 04/11/01 16:17 & S22W20 & OA  & Acc & 14000-100 & 19  & \textbf{7,2} & 991  & 810   & 1224 &      & N (S) & Dec         & 8440       & -63,73       \\
32  & 04/15/01 14:06                        & 04/15/01 14:05 & 04/18/01 00:49 & S22W85 & PH  & Dec & 14000-40  & 23  & \textbf{2,2} & 529  & 599   & 717  & Y    & N     & Dec         & 1087       & -2,75        \\
33  & 04/26/01 12:30                        & 04/26/01 12:40 & 04/28/01 05:00 & N16W29 & OA  & Acc & 5000-20   & 40  & \textbf{5}   & 1116 & 930   & 1038 & N    & Y     & Lin         & 1093       & -1,16        \\
34  & 05/03/01 09:30                        & 05/04/01 06:00 & 05/06/01 09:06 & N12W27 & 114 & Lin & 350-120   & 32  & 9,06         & 571  & 343   & 597  &      &       & Acc         & 325        & 2,11         \\
35  & 08/09/01 10:30                        & 08/10/01 01:00 & 08/12/01 11:09 & N23W13 & PH  & Acc & 200-70    & 21  & 7,61         & 1261 & 423   & 575  &      &       & Acc         & 230        & 2,60         \\
36  & 08/15/01 23:54                        & 08/16/01 00:10 & 08/17/01 11:01 &        & BA                         & Dec & 5000-100  & 20  & \textbf{7,2} & 1068 & 519   & 1209 & Y    & N     & Acc         & 819        & 5,47         \\
37  & 08/25/01 16:50                        & 08/25/01 16:50 & 08/27/01 19:38 & S18E28 & OA                         & Dec & 8000-170  & 6   & 9,1          & 1032 & 619   & 823  & Y    & N     & Dec         & 1856       & -7,39        \\
38  & 09/11/01 14:54                        & 09/12/01 10:00 & 09/13/01 02:31 & N11E22 & BA                         & Acc & 140-60    & 12  & 10,88        & 1815 & 449   & 1167 & Y    & MFR   & Acc         & 385        & 11,89        \\
39  & 09/24/01 10:30                        & 09/24/01 10:45 & 09/25/01 20:17 & S18E18 & OA                         & Acc & 7000-30   & 33  & 7,2          & 840  & 835   & 1238 &      & N     & Acc         & 759        & 7,44         \\
40  & 09/27/01 08:06                        & 09/27/01 08:15 & 09/29/01 09:29 &        & PH  & Acc & 4000-80   & 23  & \textbf{3}   & 1161 & 735   & 848  & Y    & MFR   & Acc         & 310        & 5,92         \\
41  & 10/01/01 05:30                        & 10/01/01 07:00 & 10/03/01 08:06 & S17W90 & OA                         & Acc & 1000-150  & 12  & 6,07         & 805  & 459   & 821  &      &       & Dec         & 1023       & -1,93        \\
42  & 10/09/01 11:30                        & 10/09/01 13:10 & 10/11/01 16:50 & S28E02 & OA                         & Dec & 5000-50   & 34  & 4,58         & 903  & 582   & 787  & Y    & MFR   & Acc         & 472        & 3,03         \\
43  & 10/19/01 16:50                        & 10/19/01 16:45 & 10/21/01 16:40 & N16W37 & BA  & Lin & 14000-30  & 48  & 7,93         & 982  & 636   & 877  & Y    & N     & Dec         & 1107       & -2,37        \\
44  & 10/25/01 15:26                        & 10/25/01 15:30 & 10/28/01 03:13 & S19W26 & OA                         & Lin & 14000-30  & 56  & \textbf{7,2} & 928  & 586   & 701  & N    & N     & Dec         & 1567       & -5,29        \\
45  & 11/04/01 16:35                        & 11/04/01 16:30 & 11/06/01 01:25 & N05W29 & BA                         & Dec & 14000-70  & 43  & 14,13        & 865  &       & 1266 & Y    & MFR   & Dec         & 1388       & -2,26        \\
46  & 11/17/01 05:30                        & 11/17/01 17:00 & 11/19/01 18:15 & S04E47 & OA                         & Dec & 175-90    & 11  & \textbf{7,2} & 718  & 628   & 692  & Y    & MFR   & Dec         & 2106       & -7,90        \\
47  & 11/22/01 23:30                        & 11/22/01 20:50 & 11/24/01 05:51 & S17W38 & BA                         & Dec & 8000-40   & 30  & 5,19         & 1255 & 1023  & 1374 & Y    & Y     & Acc         & 1194       & 2,66         \\
48  & 12/26/01 05:30                        & 12/26/01 05:20 & 12/29/01 05:17 & N12W71 & PH  & Dec & 14000-150 & 24  & 8,66         & 589  & 528   & 583  & N    & MFR   & Lin         & 550        & 0,14         \\
49  & 01/08/02 17:54                        & 01/08/02 18:30 & 01/10/02 15:44 &        & OA                         & Acc & 14000-90  & 29  & 6,69         & 851  &       & 907  &      &       & Acc         & 759        & 1,64         \\
50  & 01/14/02 05:35                        & 01/14/02 06:25 & 01/17/02 05:44 &        & OA                         & Acc & 12000-100 & 15  & \textbf{5}   & 618  & 277   & 576  &      &       & Lin         & 613        & -0,32        \\
51  & 04/17/02 08:26                        & 04/17/02 08:30 & 04/19/02 08:25 & S15W42 & OA                         & Dec & 5000-40   & 44  & \textbf{7,2} & 1192 & 769   & 873  & Y    & Y     & Dec         & 1819       & -7,34        \\
52  & 04/21/02 01:27                        & 04/21/02 01:30 & 04/23/02 05:00 & S14W90 & OA                         & Lin & 10000-60  & 23  & \textbf{7,2} & 811  & 639   & 818  &      &       & Lin         & 785        & 0,22         \\
53  & 05/22/02 03:50                        & 05/22/02 04:10 & 05/23/02 10:44 & S20W83 & BA                         & Dec & 500-30    & 31  & \textbf{7,2} & 1198 & 736   & 1366 & Y    & Y     & Acc         & 957        & 6,94         \\
54  & 07/15/02 21:30                        & 07/15/02 21:15 & 07/17/02 15:55 & N01W91 & PH  & Dec & 5000-175  & 8   & \textbf{7,2} & 1173 & 493   & 991  & Y    & MFR   & Lin         & 1086       & -1,37        \\
55  & 08/16/02 12:30                        & 08/16/02 12:20 & 08/18/02 18:40 & S07E10 & BA                         & Dec & 14000-60  & 33  & \textbf{7,2} & 952  & 671   & 774  & Y    & MFR   & Dec         & 2211       & -9,07        \\
56  & 09/05/02 16:54                        & 09/05/02 16:55 & 09/07/02 16:22 & N08E25 & OA                         & Acc & 14000-30  & 47  & \textbf{7,2} & 996  & 898   & 880  & Y    & MFR   & Acc         & 636        & 2,67         \\
57  & 11/09/02 13:31                        & 11/09/02 13:20 & 11/11/02 11:52 & S10W42 & OA                         & Acc & 14000-100 & 14  & 10,46        & 838  &       & 897  &      &       & Dec         & 1178       & -2,86        \\
58  & 05/28/03 00:50                        & 05/28/03 01:00 & 05/29/03 18:31 & S07W33 & BA                         & Acc & 1000-200  & 24  & \textbf{7,2} & 1149 & 906   & 999  & Y    & MFR   & Acc         & 623        & 4,72         \\
59  & 05/29/03 01:27                        & 05/29/03 01:10 & 05/30/03 16:00 & S07W46 & BA                         & Dec & 13000-200 & 7   & \textbf{7,2} & 1113 &       & 1078 & Y    & N     & Acc         & 816        & 3,42         \\
60  & 06/17/03 23:18                        & 06/17/03 22:50 & 06/20/03 08:00 & S07E44 & OA                         & Dec & 10000-200 & 7   & \textbf{7,2} & 778  &       & 733  &      &       & Lin         & 754        & -0,42        \\
61  & 10/28/03 11:30                        & 10/28/03 11:10 & 10/30/03 16:19 & S16E04 & S                          & Dec & 14000-40  & 37  & \textbf{7,2} & 568  &       & 787  & Y    & MFR   & Dec         & 8870       & -44,55       \\
62  & 11/02/03 17:30                        & 11/02/03 17:30 & 11/04/03 06:46 & S17W63 & OA                         & Dec & 12000-250 & 8   & \textbf{7,2} & 957  & 759   & 1139 &      &       & Dec         & 2725       & -15,37       \\
63  & 11/04/03 19:54                        & 11/04/03 20:00 & 11/06/03 19:19 & S17W89 & OA                         & Acc & 10000-200 & 4   & \textbf{7,2} & 936  &       & 876  &      &       & Dec         & 1108       & -2,37        \\
64  & 07/25/04 14:54                        & 07/25/04 15:00 & 07/26/04 22:25 & N08W35 & BA                         & Acc & 1000-28   & 31  & \textbf{7,2} & 1784 & 1101  & 1315 & Y    & Y     & Lin         & 1370       & -1,21        \\
65  & 09/12/04 00:36                        & 09/12/04 00:45 & 09/13/04 19:40 & N05E33 & BA                         & Acc & 14000-40  & 44  & \textbf{7,2} & 1298 &       & 965  & Y    & MFR   & Acc         & 211        & 9,64         \\
66  & 11/07/04 16:54                        & 11/07/04 16:25 & 11/09/04 09:25 & N09W08 & BA                         & Dec & 14000-60  & 28  & \textbf{7,2} & 1296 & 746   & 1032 &      & Y     & Dec         & 1666       & -6,53        \\
67  & 12/03/04 00:26                        & 12/03/04 00:07 & 12/05/04 07:00 & N09E01 & BA                         & Dec & 10000-60  & 28  & \textbf{7,2} & 727  &       & 762  &      &       & Dec         & 1596       & -5,64        \\
68  & 01/20/05 06:54                        & 01/20/05 07:15 & 01/21/05 16:45 & N14W70 & OA  & Acc & 14000-25  & 33  & \textbf{7,2} & 1894 &       & 1228 & Y    & N     & Acc         & 927        & 4,51         \\
69  & 05/13/05 17:12                        & 05/13/05 17:00 & 05/15/05 02:05 & S10W80 & BA  & Lin & 5000-40   & 33  & \textbf{7,2} & 1054 &       & 1264 & Y    & Y     & Dec         & 1878       & -8,07        \\
70  & 09/09/05 19:48                        & 09/09/05 19:45 & 09/11/05 01:00 & S09E53 & OA                         & Dec & 10000-50  & 26  & \textbf{7,2} & 1217 &       & 1423 & Y    & N     & Lin         & 1448       & -0,83        \\
71  & 09/13/05 20:00                        & 09/13/05 20:20 & 09/15/05 08:30 & S11E17 & OA                         & Acc & 1100-35   & 34  & \textbf{7,2} & 1363 &       & 1139 & Y    & MFR   & Lin         & 1222       & -1,45        \\
\bottomrule
\end{longtable}
\normalsize 
\end{center}
\end{landscape}

\begin{figure} 
 \centering
 \includegraphics[width=0.98\textwidth]{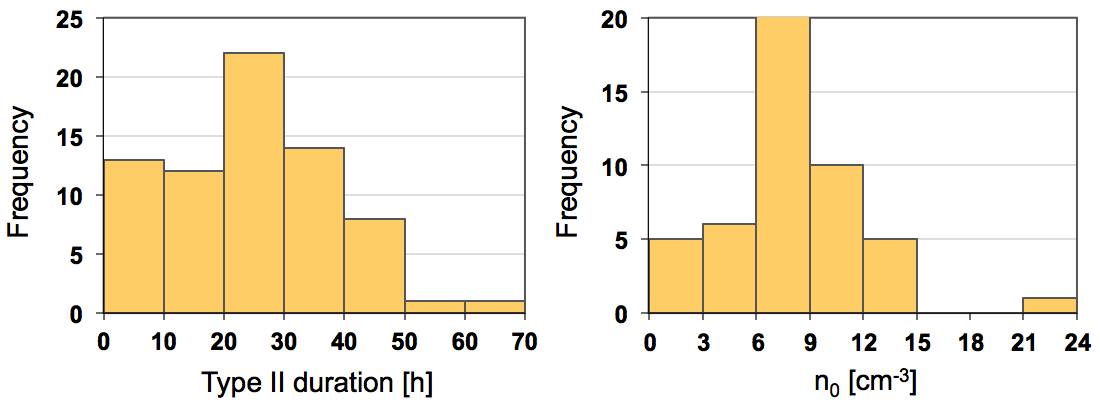}
 \caption{Left panel: Distribution of the type II emission duration for the 71 analyzed cases. The bin size is 10 h. Right panel: Distribution of the $n_0$ value used as input for the coronal/interplanetary density model. The third bin (44 events) appears truncated in the graph due to its high frequency.}
 \label{fig:density}
 \end{figure}

Column 8 of Table~\ref{tbl:associations} lists the frequency range of the 71 TII emissions here analyzed, as reported by the {\it Wind}/WAVES type II radio bursts list. All of them naturally extend down to frequencies corresponding to kilometric wavelengths. Almost 23\% of these type II events were limited to the kilometric domain, while {32\%} began in the upper detection limit of RAD2. The rest of the reported type II start frequencies were spread in the frequency detection range of the {\it Wind}/WAVES detectors. Likewise, the duration in hours of the complete TII emission as reported by the {\it Wind}/WAVES list is presented in Column 9 of Table~\ref{tbl:associations}. The type II durations, whose distribution is graphed in Fig \ref{fig:density} (left panel), range from nearly 1 h to 65 h, have an average of 25 h and a standard deviation of 13 h.  

The values of $n_0$ used as input for the coronal/interplanetary density model are presented in column 10 of Table~\ref{tbl:associations}. For 35 out of the 71 cases it was possible to deduce $n_0$ using the technique introduced in Section~\ref{s:profiles}. For 36 cases $n_0$ was user-specified, either because the results of the neural network procedure were not properly reproducing the local plasma density at the spacecraft (33 cases, typically during highly fluctuating intervals) or because there was no output from the procedure (last 3 events of the list). The value of $n_0$ manually specified (bold values in column 10 of the table) was 7.2 cm$^{-3}$ for periods of rapidly varying density, while for 7 cases the plasma line was quite stable and a better representative mean value could be adopted. The right panel of Figure \ref{fig:density} shows the distribution of $n_0$ for the 71 analyzed cases, where the third interval (6--9 cm$^{-3}$) has been truncated due to its high occurrence (44 events), in view of the fact that 29 events where manually assigned with the 7.2 cm$^{-3}$ average value. 


\begin{figure}
   \centerline{\hspace*{0.015\textwidth}
               \includegraphics[width=0.52\textwidth,clip=]{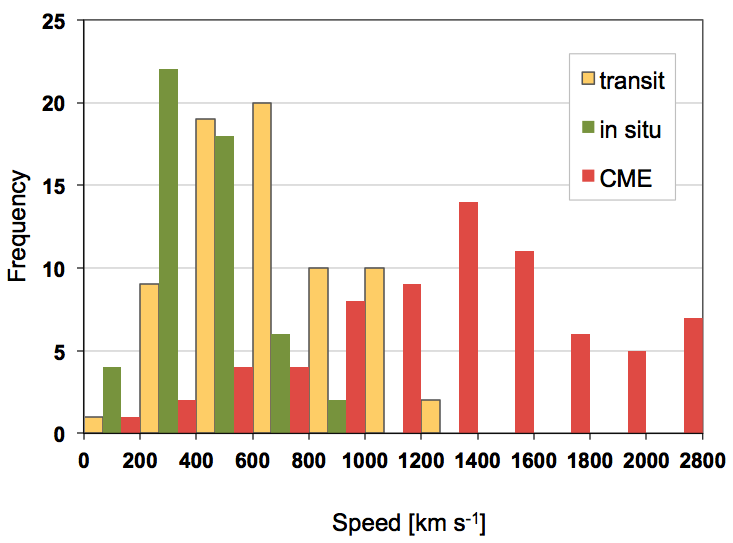}
               \hspace*{0.0\textwidth}
               \includegraphics[width=0.46\textwidth,clip=]{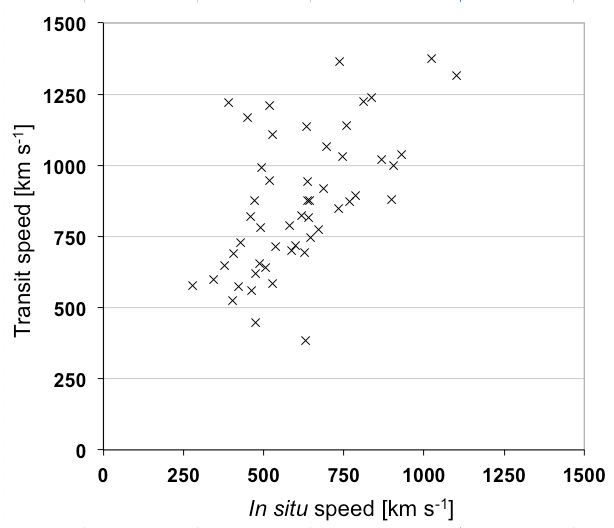}
              }
	\caption{Left panel: Speed distribution in bins of 200 km s$^{-1}$, with shock transit speed in yellow columns, \insitu shock speed in green, and CME projected speed in red columns. Right panel: Scatter plot of shock transit speeds (vertical axis) vs. \insitu shock speeds (horizontal axis) for 52 out of the 71 events.}
   \label{fig:speeds}
   \end{figure}

Speeds derived solely from the kmTII radio emission by means of the technique described in Section \ref{ss:methodology} are presented in column 11 of Table~\ref{tbl:associations}. These do not correlate well with shock speeds derived from \insitu measurements, as also noted by CSK2007, but do however with shock transit speeds. Shock transit speeds, determined from the time difference between the shock arrival time at 1 AU and the time of the first CME detection, are listed in column 13 of Table~\ref{tbl:associations}. Their distribution is presented in the left panel of Figure~\ref{fig:speeds} in yellow columns, together with that of the corresponding \insitu speeds at {\it Wind} in green columns. The latter were obtained from the shock properties at \url{http://www.cfa.harvard.edu/shocks/}, which are determined using the methods described in \citet{Pulupa-etal2010}. Values are presented in column 12 of Table~\ref{tbl:associations} and were available for 73\% of the events (52) in the mentioned data source. Average values of the shock transit and the \insitu speeds are 895 km s$^{-1}$ and 613 km s$^{-1}$ respectively, discrepancy also evident in the right panel of Figure~\ref{fig:speeds}. For comparison, in the speed histogram the CME speed in the corona is also shown (red columns), corresponding to the linear fit through the plane-of-sky projected height-time points available in the CDAW SOHO/LASCO CME Catalog. Note that there are 52 CMEs (73\%) with coronal speeds above 1000 km s$^{-1}$, and that the last bin of the histogram includes CMEs with speeds between 2000 and 2800 km s$^{-1}$.

Columns 14 and 15 of Table~\ref{tbl:associations} display ICME characteristics when available at the RC ICME list. Evidence of bidirectional suprathermal electron strahls (BDEs) in ACE observations is indicated for 43 events by ``Y'' (yes) and ``N'' (no). BDEs were identified in 39 events (91\%) out of those. This percentage contrasts with the values obtained when computing the complete RC ICME list: out of 303 ICMEs, 204 (67\%) exhibit BDEs while 99 (33\%) do not. Presence of magnetic cloud (MC) signatures is indicated in column 15, where ``Y'' means that a MC has been reported in association with the ICME or that the ICME has the clear features of a MC, ``MFR'' indicates that the ICME shows evidence of a rotation in field direction, but lacks some other characteristics of a MC, for example a smoothly rotating and enhanced magnetic field, and ``N'' means that the ICME is not a reported MC and lacks most of its typical features. The letter ``S'' in parentheses after the ``N'' stands for ``Schwenn'' and ``B'' for ``Berdichevsky {\it et al.}''; and indicates that the event was not listed in the RC ICME list, however it was listed by the author in parentheses and not recognized as a MC. Information on this matter could be found for 57 events. Fourteen events were identified as MCs in the RC ICME list, 16 more as only exhibiting magnetic field rotation, and 27 as not being a MC by various authors.

\begin{figure} 
\centering
\includegraphics[width=0.70\textwidth]{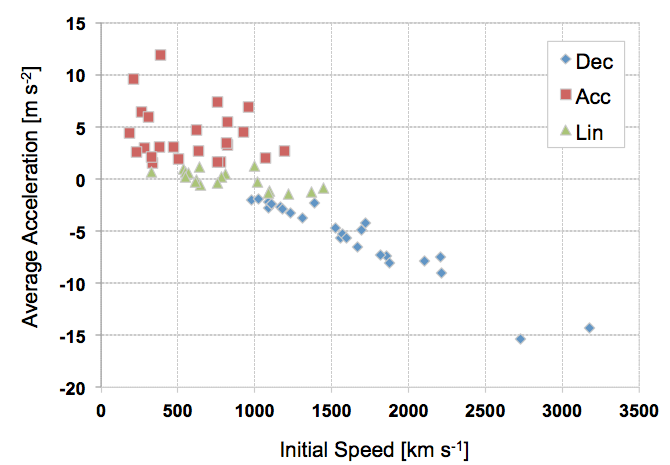}
\caption{Initial speed vs. average acceleration, both deduced from the fit connecting the CME, kmTII and shock arrival points.}
\label{fig:ViniAmed}
\end{figure}

Information on different kinematical parameters arises from the descriptive profiles of the CME/shock, i.e. the fit to the kmTII points connecting the first appearance of the CME in the coronagraph and the shock arrival at 1 AU. Some of the quantities that can be derived from the descriptive fit are the initial speed and the average acceleration, listed in columns 17 and 18 of Table \ref{tbl:associations}. The terms ``initial speed" and ``average acceleration" refer respectively to the slope of the descriptive profile for $t$=0 and the the second derivative of the distance-time expression of each event's propagation profile. Although for symplicity the first and second derivative of the descriptive profiles are here called speed and acceleration, physical interpretations should be done with care.  The initial speed does not show a preference with source region latitude, although it is notable that all CMEs with initial speed higher than 2000 km s$^{-1}$ originated in the southern hemisphere. This could be either fortuitous or due to a north-south asymmetry. Figure~\ref{fig:ViniAmed} is a scatter plot of average acceleration vs. initial speed from the descriptive profiles. Two outliers have been removed from the plot, namely those corresponding to events \#31 and \# 61, given their implausibly high initial speed. Note that corrected speed values as derived by \citet{Howard-etal2008} are 5455 km s$^{-1}$ and 7531 km s$^{-1}$ respectively for these events. In the figure, events have been classified according to their speed profile after the fitting: decelerated in blue diamonds, accelerated in red squares, and linear in green triangles. Once more, linear events are considered those with an acceleration between $\pm$1.5 m s$^{-2}$. It is straightforward that the initially fastest events are those that decelerate the most, while slower CMEs tend to accelerate. In addition, the decelerating process appears more uniform than the accelerating one, with initial speed values more spread with respect to acceleration. 

\section{SAT forecasting}

\subsection{Improved linear method}\label{s:linearpred}

The linear method here presented is a modified version of the SAT forecasting technique developed by CSK2007. The analysis is implemented, on the one hand, to assess the impact of the improvements introduced in Section \ref{ss:methodology} on the quality of the predictions, and on the other to obtain a reference for direct comparison with the results obtained in Section \ref{s:2ndopred}. This kmTII-based methodology assumes a constant speed of the MHD shock at distances corresponding to the kmTII emissions along the Sun-Earth line. Although this may seem to contradict the propagation profiles described in Section~\ref{s:profiles}, it is a good approximation at the distances at which kmTII usually occur, as suggested by reports that support the concept of acceleration-cessation distance (\eg, \opencite{Gopalswamy-etal2001b, Reiner-etal2007, Temmer-etal2011}).

The speed $v$ is calculated from the empirical expression $v=slope*a*R_0*\sqrt{n_0}$ \citep{Reiner-etal1998b} where $slope$ is the frequency drift rate of the associated kmTII emission, $a$ is a constant (9 or 18 if the radio emission occurs at the fundamental or the harmonic of the local plasma frequency respectively), $R_0=1.5 \times 10^8$ km, and $n_0$ is the electron density at Earth. Once the shock velocity is known, its distance from the Sun and arrival time can be obtained, provided that the distance of the emission along the Sun-Earth line is known. This is achieved by means of a coronal/interplanetary electron density model, in particular the \citet{Leblanc-etal1998} model, introduced in Section \ref{ss:methodology}. 

Aside from the assumption of nearly constant speed propagation at the distances associated with kmTIIs, there are others that must be taken into account. To begin with, it is assumed that the source of the radio emission propagates along the Sun-Earth line, meaning that in the case of a limb full halo CME the shock parcels responsible for the emission are those corresponding to the component travelling in the Earth's direction. It is also assumed here that events are not complex, \ie they are not interacting with other solar wind structures, and that they travel through a stationary and quiet IP medium. The pre-existing conditions of the IP medium on which CMEs/shocks propagate may modify their three-dimensional propagation, as found by several authors (\eg \opencite{Pohjolainen-etal2007, Kilpua-etal2009, Zuccarello-etal2012, Panasenco-etal2013, Lugaz-etal2014}).

As mentioned earlier, one of the main drawbacks of the CSK2007 method is the fixed value of 7.2 cm$^{-3}$ adopted for $n_0$. This value is required not only for the determination of $v$, but also by the coronal/interplanetary density model to deduce the distance at which a particular emission takes place. To overcome the limitations imposed by the use of a fixed $n_0 = 7.2$ cm$^{-3}$, new values of $n_0$ were determined by means of the procedure presented in Section \ref{ss:methodology}, whenever feasible. The other major shortcoming is the computation of the $slope$ parameter, formerly calculated as the linear interpolation between two points selected in the $1/f$ dynamic power spectrum of the emission. As explained in Section \ref{ss:methodology}, a variable number of frequency points can now be selected in the spectrum to incorporate all the ``patches'' of the usually noisy and intermittent kmTII emission. A linear fit to all the points is then applied (see white line in Figure \ref{fig:TIIemission}), obtaining this way a more robust estimation of $slope$. 

While this is a relatively simple and highly empirical technique, it has proved to be accurate and flexible enough to allow its application to emissions with spectra of several different qualities. Other techniques based in image processing or automatic spectrum analyses (\eg \opencite{GonzalezE-AguilarR2009}), obtain good results only if the the emission spectrum has a relatively well-defined and isolated shape, \ie is not contaminated by other radio phenomena. These conditions are typical for energetic events that do not overlap with other major solar phenomena occurring in the same bands (\eg type III events) and present relatively stable and homogeneous values of ambient plasma density. This scenario represents only 25\% of TII events \citep{Cane-Erickson2005}, and is not the case of the majority of the events here analyzed, since most of them took place during solar maximum.

The error in the SAT estimation is defined as the difference between the true SAT as measured by {\it Wind} at 1 AU and the predicted SAT, so that a positive (negative) error means that the event is forecasted to arrive after (before) the real SAT. The application of this technique to the 71 events in Table~\ref{tbl:associations} produces an average SAT error of $\sim$4 h, while 85\% of the events are predicted with an absolute error smaller than 6 h. This represents a significant improvement with respect to the average SAT error of 7.8 h of the original CSK2007 technique. See Table~\ref{tbl:comparison} for statistical facts, where ``Linear 2P'' stands for the original version of the technique, and ``Linear MP'' represents the multi-point improved version of the method. The performance of these techniques is discussed in Section~\ref{s:discussion}, also in comparison with that of the predictive profile method presented below.

\subsection{Predictive profile method}\label{s:2ndopred} 

The descriptive profiles introduced in Section~\ref{ss:descriptive}, which consider CME onset information, distance-time points derived from kmTII emissions, and the shock arrival at 1AU, stimulated the formulation of a predictive method. The latter results more comprehensive than the linear version presented in the previous section, because it not only considers kmTII information, but also CME data, so as to have a more complete scenario potentially helpful for forecasting. Furthermore, the linear fit through the kmTII points presented above does not appear representative of the various descriptive profiles found for the 71 events (Section~\ref{s:characteristics}). Therefore, the same simple mathematical equations \ref{eqn:acc} and \ref{eqn:dec} used to simulate the descriptive profiles introduced in Section~\ref{ss:descriptive}, are now used to fit the CME and kmTII points in a predictive fashion, \ie disregarding the shock arrival. Same assumptions done for the previous method hold: nearly Sun-Earth propagation of shock-emmiting parcel, non-complex events, and propagation through a stationary and quiet IP medium.

\begin{figure}
   \centerline{\hspace*{0.015\textwidth}
               \includegraphics[width=0.49\textwidth,clip=]{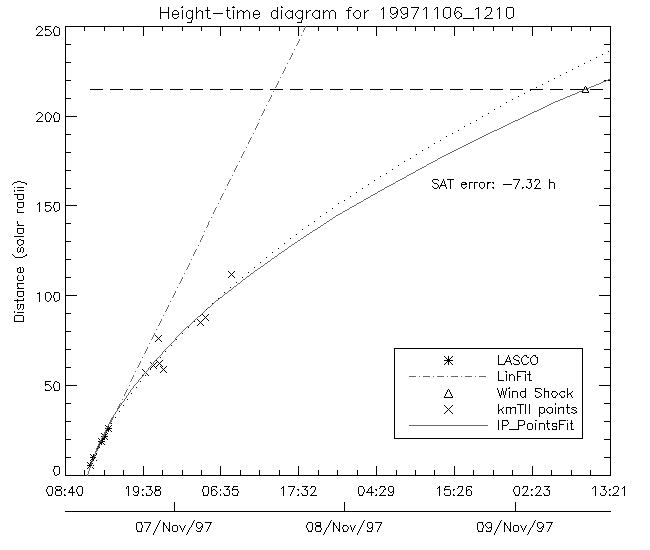}
               \hspace*{-0.03\textwidth}
               \includegraphics[width=0.49\textwidth,clip=]{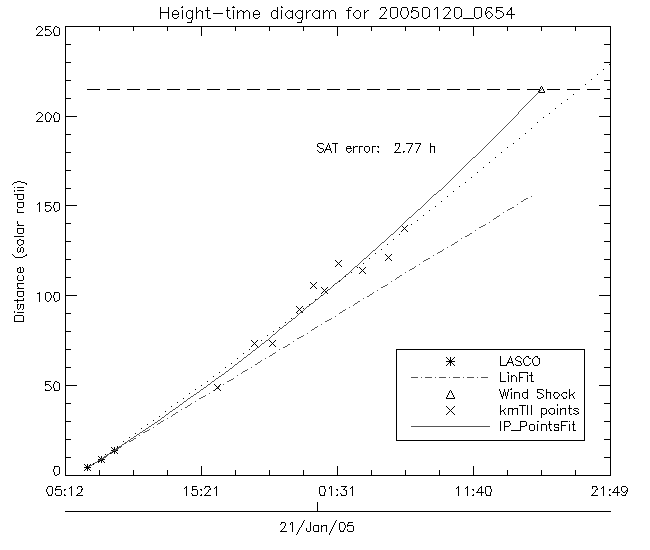}
              }
	\caption{The predictive profile method (black dotted line) in comparison with the descriptive profile (black solid line) and the linear fit through the SOHO/LASCO points (dash-dotted line), for the events starting on 6 November 1997 and 20 January 2005. The SOHO/LASCO projected height-time points are represented by asteriscs, the kmTII distance-time points by crosses, and the shock arrival at 1 AU by a triangle.}
   \label{fig:2ndpred}
   \end{figure}

Figure~\ref{fig:2ndpred} shows the various propagation profiles traced for the events starting on 6 November 1997 (left panel - decelerated case) and 20 January 2005 (right panel - accelerated case). The black solid line represents once more the descriptive profiles through the CME, kmTII, and shock arrival points; while the dash-dotted line is the linear fit through the CME projected height-time points (asterisks). The predictive profile, \ie disregarding the shock arrival data point, is represented by the black dotted line and accompanied by the corresponding SAT error in hours. Although it is straightforward that the Sun-Earth propagation of CMEs/shocks is not simple, the second-order predictive method approximates the CME and kmTII distance-time points, also by forecasting almost half of the events with a SAT error less than 6 h (see last column of Table~\ref{tbl:comparison}). The average SAT error obtained with this technique for all events yields 9.1 h, much larger than the one achieved with the improved linear method. This average error is still comparable to those of SAT predictions based on metric type II emissions, which range from $\sim$8 to 12 h (\eg, \opencite{Dryer-Smart1984, Smith-Dryer1990, Fry-etal2001}).

\begin{table}
\caption{Statistical facts on SAT errors for the CSK2007 technique (column Linear 2P), the improved linear method (Linear MP), and the predictive profile method (Pred-Profile).}
\label{tbl:comparison}
\begin{tabular}{lccc}
\toprule
 & Linear 2P & Linear MP & Pred-profile \\
\midrule
Mean absolute SAT error [h] & 7.8 & 3.8 & 9.1 \\
Median SAT error [h] & 4.3 & 2.3 & 7.3 \\
Standard deviation & 10 & 4.8 & 8.8 \\
Events with abs. SAT error $<$ 3 h [\%] & 37 & 58 & 24 \\
Events with abs. SAT error $<$ 6 h [\%] & 65 & 85 & 48 \\
Events with abs. SAT error $>$ 24 h [\%] & 7.1 & 1.4 & 7.0 \\
\bottomrule
\end{tabular}
\end{table}

\section{Summary and discussion}\label{s:discussion} 

We investigated the individual propagation of 71 ejective events after association of coronal, interplanetary, and \insitu counterparts during 1997-2007. Combining coronagraph images of CMEs, type II radio emissions in the kilometric range, and \insitu information on shocks , it is possible to build distance-time diagrams that cover the Sun-Earth journey. These give an overview of the general kinematics of CMEs and their driven shocks in the IP medium. Propagation at coronal heights is given by CME height-time information projected in the plane of the sky, while the shock arrival time at 1 AU is obtained from \insitu shock lists. The IP distance of the kmTII emissions is determined by means of a coronal/interplanetary density model, fed with a refined value of electron density at 1 AU.  

Distance-time diagrams built from CME-kmTII-shock data points were approximated by two simple fitting expressions (equations \ref{eqn:acc} and \ref{eqn:dec}), used to describe the propagation of the ejective events from Sun to Earth. These descriptive profiles can be classified according to their behaviour in decelerated (39\%), accelerated (34\%), and linear (27\%), when considering  linear events those with average acceleration within $\pm$1.5 m s$^{-2}$. In contrast, \citet{Reiner-etal2007} found few accelerated CME/shock events as evidenced by radio data. The discrepancy could be due to either a bias in the data selection procedure in one or both investigations, or incorrect CME-kmTII-shock associations, in spite of a careful examination of candidate events and double-check with other lists of CME-ICME shock associations. The propagation profiles are in agreement with the findings of \citet{Gopalswamy-etal2001b}, \citet{Reiner-etal2007}, and \citet{Liu-etal2013}. The first proposed a model based on coronagraph and \insitu data that assumes a constant acceleration from the Sun until an acceleration-cessation distance of 0.76 AU, common to all CMEs. From there onwards, a constant-speed propagation is assumed. \citet{Reiner-etal2007} investigated three stages of propagation of a set of 42 events and formulated a model that assumes a constant deceleration out to a variable heliocentric distance, followed by a constant-speed propagation at the \insitu-derived shock speed. \citet{Liu-etal2013} suggested three-propagation phases of fast events: an impulsive acceleration, a rapid deceleration, and a nearly constant speed propagation (or gradual deceleration). Previous studies solely based on coronagraph observations suggest that fast CMEs undergo an impulsive acceleration phase followed by constant or slowly decreasing speed (\eg, \opencite{Zhang-etal2001}), and found differing propagation profiles (\eg, \opencite{Vrsnak2001}), which can be regarded as indicative of different acting physical mechanisms \citep{Joshi-Srivastava2011}.

Various quantities arise from the descriptive curves of CME-kmTII-shocks propagation through the first and second derivative of the descriptive profiles. These are respectively addressed as values of speed and acceleration, though caution must be taken to avoid direct physical interpretations. Although the CME and the shock transit speed do not correlate well (as found also by \opencite{Reiner-etal2007}), the initial speed shows a strong relationship with the effective deceleration, showing that fastest CMEs decelerate at a larger rate, as previously found by several authors and in agreement with the drag force concept (\eg, \opencite{Vrsnak-Gopalswamy2002, Cargill2004, Manoharan2006, Vrsnak-etal2010, Vrsnak-etal2013}).

Characteristics of the set of analyzed events are presented, so as to typify events exhibiting kmTII emissions directly linked to shocks reaching Earth, certainly a relevant group for space weather forecasting. Nearly all CMEs related with the kmTII emissions are partial or full halo CMEs, mainly asymmetric events, in accordance with the important bias of the source region longitude toward western values. Seventy-three percent of the CMEs exhibited projected coronal speeds greater than 1000 km s$^{-1}$. For their part, type II emissions appear limited to the kilometric wavelength range (starting at $\sim$20 \Rsun) for 23\% of the cases, while 30\% began at the upper detection limit of RAD2 and extend down in frequency to few kHz (\ie to distances larger than 0.5 AU). Out of the \insitu events provided with information by the RC ICME list, a high number presented BDEs (91\%) and either MC signatures (25\%) or solely magnetic field rotation (28\%). 

The descriptive propagation profiles of CME-kmTII-shocks have the potential to help evaluate the performance of various projection-effects correction methods (Section \ref{ss:descriptive}. None of them systematicaly demonstrated to follow the descriptive profiles at coronal heights. The implementation of these methods for deprojecting coronal height-time points in views of running SAT prediction methods can have an adverse impact in the predictions, yielding errors of 12 h to 24 h. Incorrect models and/or inhomogeneities in the IP medium modifying the propagation conditions may be responsible for these errors.

In order to improve the performance of the kmTII-based SAT forecasting technique originally developed by CSK2007, a constant-speed approach, two key modifications were introduced. These were aimed at palliating the main drawbacks of that technique, concerning the determination of the crucial value of density at 1 AU required by the density model, and the selection procedure of radio emission patches in the dynamic spectrum images. In first place, instead of the $n_0$=7.2 cm$^{-3}$ used for almost all cases in the original technique, we adopted when possible the mean value of $n_0$ derived from the automatic detection of the plasma line for the day(s) when the kmTII was observed. This establishes a better constraint to the \citet{Leblanc-etal1998} density model and contributes to reduce the SAT error. However, the model sensitivity to $n_0$ established the convenience of using $n_0$=7.2 cm$^{-3}$ for events with unstable plasma frequency line, which is a common case during times of solar maximum. Issues in the estimation of CMEs/shocks location are addressed by \citet{Pohjolainen-etal2007} and include the interaction with slower CMEs and passage through a perturbed medium. Secondly, the methodology used to estimate the frequency drift rate of the kmTII emission adds the possibility to select several points on its dynamic spectrum and use a linear fit to obtain the $slope$ parameter. This is in contrast with the previous version, that was limited to the subjective selection of only two points of the kmTII emission. Type II emissions are known to exhibit a variety of behaviours that complicate their discernibility, such as multiple-lanes, differing drift rates and intermittent emission, as a consequence of coronal and interplanetary plasma structure, magnetic field topology, and relative motion of the TII source with respect to the global shock evolution \citep{Knock-Cairns2005}. The implemented improvement increases the technique accuracy when dealing with the very frequent patchy and/or noisy events, without degrading the characteristic simplicity of the original method. The reduction of the average absolute SAT error to 3.8 h is attributed to the synergystic effect of these two improvements. For a comparison, \citet{Xie-etal2006} test the performance in SAT prediction by combining input parameters of three different cone models with the ESA model \citep{Gopalswamy-etal2005b, Gopalswamy-etal2005c} and achieve average errors of almost 6 h. The statistical studies of \citet{Taktakishvili-etal2009} and \citet{Xie-etal2013}, which test the ENLIL \citep{Odstrcil-etal2005} model's predictive performance when fed with various sets of cone model parameters yielded errors ranging from 6 to 8 h.

Aside from the improved linear method of forecasting, it is attempted to employ the same expressions used for the descriptive propagation profiles but in a predictive fashion. This implies that the kmTII-derived distance-time points are only combined with the first observation of a CME in the coronagraph, while neglecting the SAT data point. In spite of being more realistic than a simple linear fit through the radio data, the average absolute error yielded by this method is comparable to the errors obtained by predictive models based on metric radio bursts \citep{Dryer-Smart1984, Smith-Dryer1990, Fry-etal2001}, which yield average errors within $\sim$8 to 14 h \citep{Fry-etal2003, McKennaL-etal2002, McKennaL-etal2006}. Average errors obtained with the presented predictive techniques may be partially explained by: (i) the linear projection to 1 AU of the kmTII points approximates better the last propagation phase of the shock wave, while if coronagraph data are introduced the prediction worsens; and (ii) although the best efforts were undertaken to find the most suitable association between each kmTII event and the CME that originated it, discrepancies found in some few cases would indicate that both phenomena are not related. Because of the latter, events with average absolute SAT error greater than 6 h for the predictive profile method reach 52\%, while for the improved linear method they account for only 15\%. 

The shortcomings of these kmTII-based forecasting techniques include: (a) the need of space-based instrumentation capable of performing low-frequency radio measurements; (b) the shorter anticipation of the forecasts, given that the kmTII emissions take place at larger distances from the Sun (from $~$30 \Rsun onwards), (c) the lack of information on which portion of the three-dimensional shock is actually producing the emission, and (d) the low amount of shock waves arriving at 1AU that are effectively associated with kmTII emissions (28\% for the investigated period). In this regard, it must be noted that there may have been more kmTII emissions than actually reported in the {\it Wind}/WAVES Type II radio bursts list, because they remained hidden in the RAD1 dynamic spectra due to its low frequency resolution close to its lower detection limit. 

Furthermore, these techniques also rely on the assumptions of non-complex events and propagation through a stationary and quiet IP medium. Therefore, their performance may be affected under the situation of interaction with other solar wind structures and deviation from a stationary propagation. Nonetheless, more than half of the events here analyzed took place during the maximum of solar cycle 23, when complex events and highly disorganized IP conditions dominated the background IP medium, and still resulting in the reduction of the SAT error of the improved linear method.

The present study appears promising towards achieving better-educated SAT forecasts, solely based on spacecraft data located in the Sun-Earth line, thus emphasizing the need of continuity of space missions that monitor wavelength ranges filtered out by the Earth's atmosphere. Although the success of the predictive profile method is relative, the descriptive profiles built for 71 Earth-directed events provide some insights on their Sun-Earth propagation up to 1 AU. The improved linear method, only based on kmTII emissions, yields however considerable improvements in SAT predictions of (I)CME-driven shocks.

%

\begin{acks}
HC and FAI acknowledge the funding of project CONVPICTPROM01 2007 of Universidad Tecnol\'ogica Nacional - Facultad Regional Mendoza. HC is member of the Carrera del Investigador Cient\'ifico (CONICET). HX, OCS and NG gratefully acknowledge support from NASA's LWS TR\&T program through grant 8-LWSTRT08-0029. The authors thank Ernesto Aguilar-Rodr\'iguez and Craig Markwardt for their support on SolarSoft and IDL routines, Cathie Meetre for guidance on neural network usage for automated density detection, Dibyendu Nandy for providing the full list of CME events corrected by the HNK method, and Ivo Dohmen for disinterested help, as well as the reviewer for helpful comments and suggestions.

The SOHO/LASCO data are produced by an international consortium of the NRL (USA), MPI f\"ur Aeronomie (Germany), Laboratoire d'Astronomie (France), and the University of Birmingham (UK). SOHO is a project of international cooperation between ESA and NASA. The STEREO/SECCHI project is an international consortium of the NRL, LMSAL and NASA/GSFC (USA), RAL and Univ. Bham (UK), MPS (Germany), CSL (Belgium), IOTA and IAS (France). SDO/AIA and SDO/HMI data are courtesy of the NASA/SDO and the AIA and HMI Science Teams.
This paper uses data from the SOHO/LASCO CME catalog generated and maintained at the CDAW Data Center by NASA and the CUA in cooperation with NRL, from the CACTus CME catalog generated and maintained by the SIDC at the ROB, and the SEEDS project supported by the NASA/LWS and AISRP programs.

\end{acks}

%
%
\bibliographystyle{spr-mp-sola}
\bibliography{Cremades-etal}

\end{article} 
\end{document}